\documentclass{article}
\usepackage{spconf}
\usepackage{amsmath,amssymb,amsfonts}
\usepackage{cite}
\usepackage{graphicx}
\usepackage{textcomp}
\usepackage{xcolor}
\usepackage{placeins}
\usepackage{booktabs}
\usepackage{array}
\usepackage{multirow}
\usepackage{tabularray}
\usepackage{url}
\usepackage{hyperref}
\hypersetup{hidelinks}

\title{S2SND-v2: Unified Online and Offline Speaker Diarization with Speaker-adapted w2v-BERT 2.0}

\name{Jincheng Wang$^{1,2}$, Li Li$^{1,2}$, Weixin Zhu$^{3}$, Yannan Wang$^{3}$, Tao Yu$^{3}$, Juan Liu$^{1,*}$, Ming Li$^{2,*}$\thanks{Corresponding Authors: Juan Liu and Ming Li.}}
\address{$^{1}$School of Artificial Intelligence, Wuhan University, Wuhan, China\\
$^{2}$School of Artificial Intelligence, the Chinese University of Hong Kong, Shenzhen, China\\
$^{3}$Tencent Ethereal Audio Lab, Tencent, Shenzhen, China\\
\texttt{liujuan@whu.edu.cn}, \texttt{mingli369@cuhk.edu.cn}}

\begin{document}
\ninept
\maketitle

\begin{abstract}
This paper presents S2SND-v2, a unified online and offline speaker
diarization framework built on S2SND that improves acoustic fea-
ture extraction and the use of the resulting representations. For
acoustic feature extraction, we introduce a speaker-adapted w2v-
BERT 2.0 front end and compare full and pruned construction
routes. To better use these features, overlap-aware activity-query
masking improves speaker embedding extraction under overlap,
while late-speaker training aligns unknown-speaker training with
online inference conditions. Combining these improvements, the
pruned system achieves competitive average diarization error rates
of 14.52\% online and 12.03\% offline across eight test sets.
\end{abstract}


\begin{figure*}[!t]
  \centering
  \includegraphics[width=0.85\textwidth]{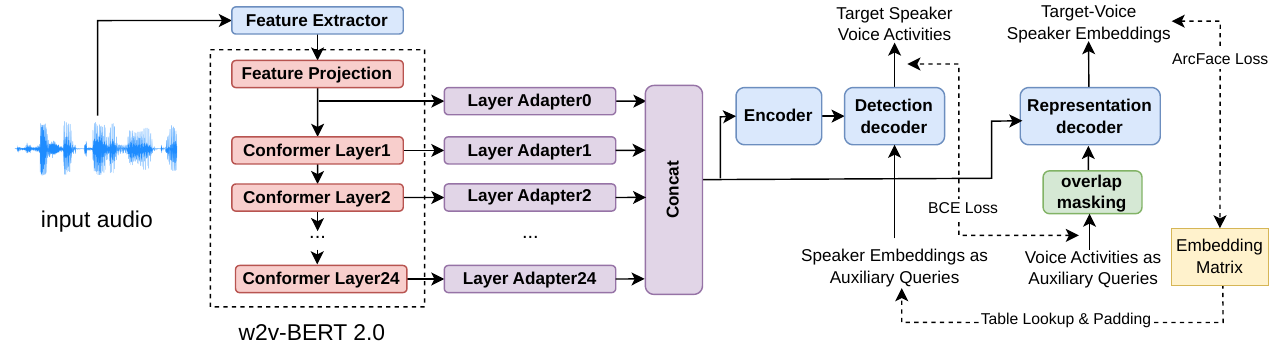}
  \caption{S2SND-v2 model framework.}
  \label{fig:training-architecture}
\end{figure*}

\section{Introduction}
\label{sec:introduction}

Speaker diarization determines who spoke when in a recording and is a
core component of meeting transcription. Early diarization systems
mainly relied on cascaded pipelines comprising speech activity
detection, segmentation, speaker embedding extraction, and clustering.
More recently, end-to-end neural diarization models have directly
predicted frame-level speaker activities, while hybrid methods have
combined local neural predictions with embedding-based clustering for
long-form recordings~\cite{fujita2019eendpit,horiguchi2020eda,kinoshita2021eendvc}.
Sequence-to-sequence prediction has also been introduced for target-speaker
voice activity detection~\cite{cheng2023target}.
For online diarization, additional mechanisms such as speaker-tracing
buffers and global/local attractors have been introduced to maintain consistent 
speaker identities across successive chunks~\cite{xue2021stb,horiguchi2023gla}. Although these advances have
substantially improved diarization, many existing methods are still
designed primarily for either offline or online use. A unified framework
that supports both modes is therefore attractive in practice.

A recent development in this direction is sequence-to-sequence
neural diarization (S2SND), which couples a detection decoder and a
representation decoder within a single framework for online and offline
inference~\cite{cheng2025s2snd}. In S2SND, speaker embeddings serve as
queries for activity prediction, while activity queries guide speaker
embedding extraction, allowing activity
estimation and speaker representation learning to reinforce each other.
Embeddings accumulated during the online pass are then reused for later
blocks and for offline re-decoding. However, its overall
performance still depends on two factors: the quality of the acoustic
representations provided to the back end, and how effectively the back
end extracts and uses speaker information under realistic online
conditions.

One direction for further improving S2SND is to strengthen the acoustic representations provided 
by its front end. The original S2SND uses a ResNet front end, whereas
self-supervised learning (SSL) has recently provided stronger and more
transferable speech representations for speech and speaker
tasks~\cite{chen2022wavlm}. This trend has also reached diarization,
where SSL encoders such as WavLM have been adopted in hybrid and neural
systems~\cite{delcroix2023msvbx,liao2026dsecbm}, and
recent work has further adapted and compressed SSL encoders for speaker
segmentation and verification~\cite{baroudi2024specializing,han2025fine,han2025efficient,li2026enhancing}.
In particular, Li \emph{et al.} combine w2v-BERT 2.0 with layer
adapters, multi-scale feature aggregation, low-rank adaptation (LoRA),
and distillation-guided structured pruning, achieving strong speaker
verification performance with both full and pruned
models~\cite{li2026enhancing}. These results motivate revisiting the
front end of S2SND with a stronger speaker-adapted SSL model that
preserves frame-level temporal resolution for diarization.

Another direction is to improve how S2SND uses acoustic
representations through query construction and training.
Although S2SND uses non-overlapping speech duration to select and weight
accepted embeddings, the activity queries used by its representation
decoder can still contain overlapping frames~\cite{cheng2025s2snd}.
Such overlap-contaminated queries may degrade target-speaker embedding
extraction. In addition, the random speaker masking strategy used for
unknown-speaker training does not reflect how unknown speakers are often
encountered in online inference. In sliding-window processing, a new
speaker is more likely to first become active in the latter part of the current input block,
creating a mismatch between random masking during training and
late-speaker conditions at test time. Together, these issues motivate improving S2SND through 
not only stronger acoustic representations but also cleaner embedding extraction and 
better-aligned unknown-speaker training.

In this work, we present S2SND-v2 and evaluate it on eight test sets. Our contributions are summarized as follows:
\begin{itemize}
  \item We integrate speaker-adapted w2v-BERT 2.0 into S2SND and compare
  full and pruned front end construction routes for unified online and
  offline diarization.
  \item We introduce overlap-aware activity-query masking to suppress
  interference from concurrent speakers in the queries used by the
  representation decoder, improving activity-conditioned speaker
  embedding extraction.
  \item We introduce late-speaker training that combines late-onset
  mixture generation with onset-aware masked-speaker selection, increasing
  training exposure to unknown speakers with late activity onsets.
\end{itemize}

\section{Methods}
\label{sec:method}

\textbf{S2SND:}
The original S2SND framework~\cite{cheng2025s2snd} unifies online and
offline diarization through two coupled decoders. An acoustic extractor
produces frame-level features, and an encoder adds temporal context.
The detection decoder uses speaker embeddings as queries to predict
activities from the encoder output, while the representation decoder
uses activity queries to extract speaker embeddings from the extractor
features. During training, speaker queries are formed through
embedding-table lookup and padding, while reference activities guide
embedding extraction. During online inference, input blocks are processed
with a sliding window. Predicted activities guide embedding extraction,
and the extracted embeddings are accumulated and reused as queries for
later blocks. Offline inference re-decodes the recording using the
accumulated embeddings.

\textbf{S2SND-v2:}
Fig.~\ref{fig:training-architecture} illustrates the proposed S2SND-v2
architecture, which retains the coupled decoders of S2SND and introduces
a speaker-adapted w2v-BERT 2.0 front end. Building on this architecture,
we also refine query construction and training to better use the acoustic
representations. Specifically, overlap-aware activity-query masking
supports speaker embedding extraction under overlapping speech, while
late-speaker training better matches the conditions under which unknown
speakers appear during online inference. These changes preserve the
unified online and offline inference framework. The following subsections
describe the front end integration and the two back end modifications.

\subsection{Speaker-adapted w2v-BERT 2.0 Front End}
\label{sec:frontend}

The original S2SND uses a ResNet front end. We instead introduce the speaker-adapted w2v-BERT 2.0
models of Li \emph{et al.}~\cite{li2026enhancing} to provide stronger
speaker-verification frame-level representations. To integrate them
into S2SND, we retain the layer-adapter and multi-scale feature
aggregation (MFA) design of the speaker-verification model, but remove
utterance-level pooling and the verification embedding head. MFA combines
representations across encoder depths into a frame-level feature sequence,
which preserves the temporal resolution required for diarization. The
resulting sequence is then fed to the original S2SND back end: one
path provides contextualized features for the detection decoder, while
the other provides memory features for the representation decoder.

\textbf{Full and pruned adaptation.} 
For the full w2v-BERT front end, we adapt the pretrained encoder with LoRA while keeping the S2SND 
back end trainable. To reduce computational cost while retaining the benefits of these acoustic 
representations, we also investigate a compact front end obtained through distillation-guided structured 
pruning. The pruned encoder uses the same frame-level feature interface but is fully fine-tuned 
without LoRA during joint training. Both configurations provide frame-level acoustic 
representations to the same S2SND back end.

\begin{figure}[!t]
  \centering
  \includegraphics[width=0.60\columnwidth]{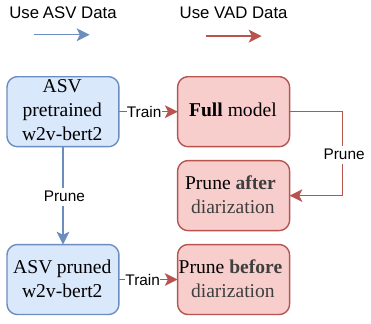}
  \caption{Three w2v-BERT front end construction routes. Blue arrows denote
  training with speaker-verification data; red arrows denote training with
  diarization data.}
  \label{fig:frontend-routes}
\end{figure}

\textbf{Front end construction routes.}
We study three front end construction routes, shown in
Fig.~\ref{fig:frontend-routes}. The \emph{full w2v-BERT} route directly
adapts the full speaker-verification model to diarization. The
\emph{prune-before-diarization} route starts from a model already pruned 
for speaker verification and then adapts it to
diarization. The \emph{prune-after-diarization} route first adapts the
full front end to diarization, then prunes it with diarization data, and
finally fine-tunes the resulting pruned system. We compare these 
construction strategies rather than isolate the effect of pruning.
For the prune-after-diarization route, pruning is guided by distillation
from the diarization-adapted full model. The student is trained to match
the hidden representations of the teacher model with the feature-distillation loss
\begin{equation}
\mathcal{L}_{\mathrm{feat}}=
\frac{1}{KT}\sum_{\ell=0}^{K-1}\sum_{t=1}^{T}
\left[\frac{\|\Delta_{\ell,t}\|_1}{d}
-\cos(\mathbf{h}^{T}_{\ell,t},\mathbf{h}^{S}_{\ell,t})\right],
\label{eq:distillation}
\end{equation}
where $\Delta_{\ell,t}=\mathbf{h}^{T}_{\ell,t}-\mathbf{h}^{S}_{\ell,t}$,
$T$ is the sequence length, $d$ is the hidden-state dimension, and
$K=25$ comprises the feature-projection output and 24 Conformer-layer
outputs adapted by the layer adapters. Superscripts
$T$ and $S$ denote teacher and student, respectively. During pruning,
this distillation objective is combined with sparsity constraints over
feed-forward dimensions, convolution channels, and attention heads.
Removing the selected structures reduces the w2v-BERT encoder
from 580M to 91M parameters, excluding the layer adapters and the S2SND
modules. Optimization and training schedules are given in
Section~\ref{sec:training}.

\subsection{Overlap-aware Activity-query Masking}
\label{sec:overlap}

Besides improving the acoustic front end, S2SND-v2 also refines how
speaker information is used in the original S2SND back end. Our first
back end modification focuses on the quality of speaker embeddings. In S2SND, accepted embeddings are selected and
weighted by their non-overlapping speech duration~\cite{cheng2025s2snd}.
However, the activity queries sent to the representation decoder may
still contain overlapping frames. As a result, overlap is considered
during embedding selection, but it can still affect the earlier
embedding-extraction stage. The representation decoder may therefore
attend to regions containing speech from multiple speakers.

We set activity-query values to zero at frames with two or more active speakers before passing the queries to the representation decoder. The mask uses reference activities during training and predicted activities during inference. During training, we assign the extracted embedding to the reserved class 0 under ArcFace supervision when the target speaker’s cumulative non-overlapping speech duration within the current input block is below 1.28s, since embeddings derived from insufficient speech may be unreliable. The same duration threshold governs whether an embedding is accepted into the speaker embedding matrix during inference. Together, these changes support speaker embedding extraction from cleaner evidence for subsequent reuse in subsequent inference windows.

\subsection{Late-speaker Training}
\label{sec:late}

Our second back end modification targets how the
pseudo-speaker query is trained for unknown-speaker detection. In the
original S2SND training procedure, one speaker's embedding is randomly masked,
and that speaker's activity target is assigned to the pseudo-speaker
query~\cite{cheng2025s2snd}. This design enables unknown-speaker
detection, but it does not fully match online inference. In
sliding-window decoding, a new speaker is more likely to first become
active in the latter part of the current input block, while the earlier
part is dominated by already-known speakers. Random masking therefore
teaches the model to detect an arbitrary speaker whose embedding has been masked, rather than a
newly arriving speaker with late activity onset.

To better match this online condition, we combine late-onset mixture
generation with onset-aware masked-speaker selection. Simulated mixtures
with at least two speakers use late-onset generation with probability
0.7, placing one speaker's onset uniformly between 50\% and 80\% of the
mixture duration. For both real and simulated training windows, we uniformly
select a speaker first active in the latter half with probability 0.7;
otherwise, or if no such speaker exists, we sample uniformly from all active speakers.
Late-onset mixtures always enable masking; other training windows retain the
original masking schedule. Generation and selection use independent
random draws. This increases exposure to late-arriving unknown speakers
while leaving inference unchanged.

\section{Experimental Setup and Results}
\label{sec:experiments}
\begin{table*}[!t]
  \centering
  \caption{Front end routes and cumulative S2SND-v2 ablation: DER (\%)
  with a 0s collar. All models use four simulation-source corpora and eight
  real-data training splits. Parameters are shown as w2v-BERT encoder+remaining modules
  (millions). FLOPs measure total computation of the full system when decoding
  an 8s window once. Additions build on prune-before-diarization. Bold indicates
  the best displayed result per dataset and mode, including ties; lower is better.}
  \label{tab:ablation}
  \footnotesize
  \newcommand{\bestval}[1]{\textbf{#1}}
  \newcommand{\tablehead}[1]{\textbf{#1}}
  \noindent
  \begin{tblr}{
    width=\textwidth,
    colspec={Q[l,wd=0.185\textwidth] Q[c,wd=0.065\textwidth] Q[c,wd=0.065\textwidth] Q[c,wd=0.045\textwidth] *{9}{X[c]}},
    colsep=2.0pt,
    rowsep=0.35pt,
    vline{2,5,13}={0.35pt},
    hline{1,2,4,6,12}={0.35pt},
    hline{8,10}={1}{0.35pt,dashed},
    hline{8,10}={2-13}{0.35pt},
  }
    \tablehead{Configuration} & \tablehead{Param. (M)} & \tablehead{FLOPs (G)} & \tablehead{Mode}
      & \tablehead{AMI-SDM}
      & \tablehead{AIS-4} & \tablehead{AliM} & \tablehead{NSF}
      & \tablehead{MSD} & \tablehead{DH3} & \tablehead{RAMC}
      & \tablehead{VoxC} & \tablehead{Avg.} \\
    \SetCell[r=2]{l} Full w2v-BERT
      & \SetCell[r=2]{c} 580+33 & \SetCell[r=2]{c} 475.37 & Online  & \bestval{16.92} & \bestval{10.78} & \bestval{11.88} & 21.22 & \bestval{17.77}
      & 16.49 & 10.96 & 11.92 & 14.74 \\
      & & & Offline & \bestval{13.27} & \bestval{9.66} & \bestval{9.90} & 15.47 & \bestval{13.58}
      & 15.09 & \bestval{10.43} & 9.80 & 12.15 \\
    \SetCell[r=2]{l} Prune-after-diarization
      & \SetCell[r=2]{c} 91+33 & \SetCell[r=2]{c} 75.63 & Online  & 17.76 & 12.12 & 12.14 & 23.36 & 17.80
      & 15.79 & 11.19 & 12.27 & 15.30 \\
      & & & Offline & 13.52 & 10.54 & 10.02 & 17.66 & 13.62
      & 14.48 & 10.56 & 10.48 & 12.61 \\
    \SetCell[r=2]{l} Prune-before-diarization
      & \SetCell[r=2]{c} 91+33 & \SetCell[r=2]{c} 75.63 & Online  & 18.05 & 10.98 & 12.14 & 24.36 & 17.86
      & \bestval{15.68} & 11.06 & 11.96 & 15.26 \\
      & & & Offline & 14.41 & 9.93 & 10.17 & 18.62 & 13.73
      & \bestval{14.06} & 10.50 & 10.14 & 12.70 \\
    \SetCell[r=2]{l} \hspace{0.6em}+ Overlap masking
      & \SetCell[r=2]{c} 91+33 & \SetCell[r=2]{c} 75.63 & Online  & 17.30 & 11.18 & 11.98 & 20.94
      & 18.27 & 16.37 & 11.09 & \bestval{10.84} & 14.75 \\
      & & & Offline & 13.64 & 9.93 & 10.08 & 15.47
      & 13.91 & 15.23 & 10.61 & \bestval{8.97} & 12.23 \\
    \SetCell[r=2]{l} \hspace{0.6em}++ Late-speaker training
      & \SetCell[r=2]{c} 91+33 & \SetCell[r=2]{c} 75.63 & Online  & 16.95 & 11.21 & 12.20 & \bestval{20.23}
      & 17.91 & 15.70 & \bestval{10.92} & 11.04 & \bestval{14.52} \\
      & & & Offline & 13.34 & 9.97 & 10.08
      & \bestval{15.06} & 13.77 & 14.45 & 10.51 & 9.07
      & \bestval{12.03} \\
  \end{tblr}
  \vspace{-2pt}
  \begin{minipage}{0.96\textwidth}
    \scriptsize AIS-4: AISHELL-4; AliM: AliMeeting; NSF: NOTSOFAR-1;
    MSD: MSDWild; DH3: DIHARD III; VoxC: VoxConverse.
  \end{minipage}
\end{table*}

\begin{table*}[!t]
  \centering
  \caption{DER (\%) comparison on eight test sets without oracle VAD and with
  a 0s collar. Online and offline systems are grouped separately; dashes
  denote unavailable results. Avg. is omitted for incomplete coverage.
  Both S2SND-v2 variants are trained in this work and include overlap-aware
  activity-query masking and late-speaker training.
 External results are from the cited publications or official repositories; evaluation protocols may differ.
  Bold values are the best within each
  mode based on displayed values, including ties; lower is better.}
  \label{tab:comparison}
  \footnotesize
  \newcommand{\bestval}[1]{\textbf{#1}}
  \newcommand{\tablehead}[1]{\textbf{#1}}
  \noindent
  \begin{tblr}{
    width=\textwidth,
    colspec={Q[l,wd=0.20\textwidth] *{9}{X[c]}},
    colsep=3.2pt,
    rowsep=0.35pt,
    vline{2,10}={0.35pt},
    hline{1,2,4,5,6,8,9,10,11,12,13}={0.35pt},
  }
    \tablehead{System} & \tablehead{AMI-SDM} & \tablehead{AIS-4}
      & \tablehead{AliM} & \tablehead{NSF} & \tablehead{MSD}
      & \tablehead{DH3} & \tablehead{RAMC} & \tablehead{VoxC}
      & \tablehead{Avg.} \\
    \tablehead{Online inference} & & & & & & & & & \\
    S2SND (retrained)~\cite{cheng2025s2snd}
      & 19.48 & 12.69 & 13.71 & 26.87 & 20.34 & 19.21 & 11.31 & 13.62 & 17.15 \\
    S2SND-v2 (ReDimNetV2-B6)
      & 18.92 & 11.32 & 14.01 & 24.88 & 19.98 & 18.25 & 11.54 & 11.70 & 16.33 \\
    \textbf{S2SND-v2 (w2v-BERT 2.0)}
      & \bestval{16.95} & \bestval{11.21} & \bestval{12.20} & \bestval{20.23} & \bestval{17.91}
      & \bestval{15.70} & \bestval{10.92} & \bestval{11.04} & \bestval{14.52} \\
    \tablehead{Offline inference} & & & & & & & & & \\
    Pyannote v3.1~\cite{bredin2023pyannote}
      & 22.40 & 12.20 & 24.40 & -- & 25.30 & 21.70 & 22.20 & 11.30 & -- \\
    DSE-CBM~\cite{liao2026dsecbm}
      & 16.70 & \bestval{9.80} & 14.90 & -- & 19.20 & -- & 10.90 & \bestval{8.60} & -- \\
    DiariZen-Large-s80-v2~\cite{palka2026vbx}
      & 13.90 & \bestval{9.80} & 10.80 & 16.70 & 15.60 & 14.50 & 11.00 & 9.10 & 12.67 \\
    S2SND (retrained)~\cite{cheng2025s2snd}
      & 16.09 & 10.68 & 11.35 & 20.77 & 16.58 & 17.58 & 10.58 & 11.99 & 14.45 \\
    S2SND-v2 (ReDimNetV2-B6)
      & 15.74 & 10.02 & 11.91 & 20.08 & 16.04 & 17.10 & 10.89 & 10.07 & 13.98 \\
    \textbf{S2SND-v2 (w2v-BERT 2.0)}
      & \bestval{13.34} & 9.97 & \bestval{10.08} & \bestval{15.06} & \bestval{13.77}
      & \bestval{14.45} & \bestval{10.51} & 9.07 & \bestval{12.03} \\
  \end{tblr}
\end{table*}

\subsection{Datasets}
\label{sec:datasets}

We generate simulated mixtures from 3D-Speaker~\cite{zheng2023threedspeaker},
KeSpeech~\cite{tang2021kespeech}, VoxCeleb2~\cite{chung2018voxceleb2},
and VoxBlink2~\cite{lin2024voxblink2}. Real-data adaptation and evaluation
use eight diarization benchmarks: AMI-SDM~\cite{carletta2005ami},
AISHELL-4~\cite{fu2021aishell4}, AliMeeting~\cite{yu2022m2met},
NOTSOFAR-1~\cite{vinnikov2024notsofar}, MSDWild~\cite{liu2022msdwild},
DIHARD III~\cite{ryant2021dihard3}, MagicData-RAMC~\cite{yang2022ramc},
and VoxConverse~\cite{chung2020voxconverse}. We use their training splits
for adaptation and follow the test splits described by Han \emph{et al.}~\cite{han2025efficient}.
Only the first channel is used for multichannel recordings. DER is computed
without oracle VAD and with a 0s collar; Avg. denotes the unweighted eight-set mean.

\subsection{Experimental Setup}
\label{sec:training}

We initialize w2v-BERT from speaker-verification models and follow the
adaptation and pruning framework of Li \emph{et al.}~\cite{li2026enhancing}.
Training consists of simulation-only training followed by two mixed-data
stages with a 4:1 simulated-to-real ratio. We train on 8s windows using AdamW 
with frame-level binary cross-entropy and ArcFace-based speaker classification losses.
The learning rate is $10^{-4}$ in the first two stages and $10^{-5}$ in the last stage.

For the full w2v-BERT front end, simulation-only training first freezes the extractor
and trains the diarization back end, then enables LoRA and layer adapters
while keeping pretrained encoder weights fixed. LoRA adapters with rank
64 and scaling factor $\alpha=128$ (no dropout) are applied to the query,
key, value, and output projections of the self-attention modules and to
the two feed-forward dense layers of each Conformer block. This
configuration continues in the mixed-data stages. For
prune-before-diarization, the pruned encoder is frozen in the
simulation-only stage and fully updated in the mixed-data stages,
without LoRA. Both configurations use the same feature interface described in Section~\ref{sec:frontend}.
For prune-after-diarization, the full diarization-adapted encoder serves
as a frozen teacher during training only. Student and sparsity-variable
learning rates are $2\times10^{-4}$ and $2\times10^{-2}$, respectively.
Sparsity increases to a target of 85\% over 10,000 updates, 
matching the sparsity of the model pruned for speaker verification.
We match 25 teacher hidden-state outputs for 20 epochs, remove the
selected structures, and distill for another 20 epochs without sparsity
regularization before the two mixed-data diarization stages.

We retrain the ResNet-based S2SND baseline using the same data and
evaluate the back end modifications on prune-before-diarization.
We also train S2SND-v2 with ReDimNetV2-B6 as front end~\cite{yakovlev2026redimnet2},
keeping both back end refinements, training data, schedules, and decoding
settings unchanged.
For online inference, all S2SND variants use an 8s sliding window 
with a step of 0.64s and a look-ahead of 0.16s, giving a nominal 
output delay of 0.80s. Offline inference re-decodes the recording 
using accumulated speaker embeddings, with the same model and decoding 
hyperparameters. Other decoding settings follow S2SND.

\subsection{Overall Results}
\label{sec:results}

\noindent\textbf{Front end construction routes.}
Table~\ref{tab:ablation} compares the three w2v-BERT front end construction routes.
Among these routes, the full
w2v-BERT model achieves the lowest average DER, reaching 14.74\%
online and 12.15\% offline. The two pruning routes are close to each
other: prune-before-diarization obtains 15.26\% and 12.70\%, while
prune-after-diarization gives 15.30\% and 12.61\%. Thus, pruning the
w2v-BERT encoder from 580M to 91M parameters incurs a modest increase
in average DER while reducing per-window computation
from 475.37G to 75.63G FLOPs.
Since neither pruning order consistently
outperforms the other across datasets or inference modes, we use
prune-before-diarization in the remaining experiments. This route
directly reuses a pruned speaker-verification model and avoids
training a full diarization teacher before pruning.

\noindent\textbf{Back end refinements.}
We next evaluate overlap-aware activity-query masking and late-speaker
training on top of the prune-before-diarization front end. To assess
their cumulative contribution, we introduce them sequentially while
keeping the front end configuration and training data unchanged.
The first refinement combines overlap-aware activity-query masking with 
duration-conditioned identity supervision, 
reducing average DER from 15.26\% to 14.75\% online and from 12.70\% to 12.23\% offline. 
The gain is not uniform across datasets, with a few sets showing
slight degradation; this may reflect insufficient activity-query information
after masking when non-overlapping speech is limited.
Adding late-speaker training further reduces average DER to
14.52\% online and 12.03\% offline, improving five of the
eight test sets in each mode relative to the preceding configuration.
These gains may also carry over to offline re-decoding, as it reuses
speaker embeddings accumulated during online inference. Taken together,
these results show that the back end refinements
complement the stronger front end rather than replace its effect.
With the pruned 91M w2v front end, the complete system achieves
average DER comparable to, and slightly lower than, the full 580M
configuration without back end refinements.

\subsection{Comparison with Existing Systems}
\label{sec:comparison}

Table~\ref{tab:comparison} compares the complete S2SND-v2 system with
our retrained S2SND baseline and published diarization systems.
Under the same inference strategy and a 0.80s output
latency, the complete system lowers the online average DER from 17.15\%
to 14.52\% relative to our
retrained S2SND baseline, and lowers the offline average DER from 14.45\%
to 12.03\%, improving performance on all eight test sets in both modes.
With the same back end refinements and training and decoding settings,
the pruned w2v-BERT 2.0 variant outperforms the ReDimNetV2-B6 variant
on all eight test sets in both modes, with average DERs of 14.52\%
versus 16.33\% online and 12.03\% versus 13.98\% offline. This comparison
further supports the choice of speaker-adapted w2v-BERT 2.0 as the
acoustic front end.

Beyond the improvements over the retrained S2SND baseline, S2SND-v2
also achieves a lower offline average DER than DiariZen-Large-s80-v2
(12.03\% vs.\ 12.67\%), with the lowest DER on six datasets, while
DSE-CBM performs best on VoxConverse.

\section{Conclusion}
\label{sec:conclusion}

This paper presented S2SND-v2, combining a speaker-adapted w2v-BERT 2.0 front 
end with improved query construction and training. Structured pruning reduces 
w2v-BERT encoder parameters by approximately 85\% while preserving most of the full-model gains. 
Overlap-aware activity-query masking and late-speaker training further improve the pruned system, 
achieving average DERs of 14.52\% online and 12.03\% offline across eight test sets. The resulting 
pruned system outperforms S2SND in both modes, achieving competitive online performance and the 
lowest reported offline average DER among the listed systems.

\clearpage
\begingroup
\setlength{\itemsep}{0pt}
\setlength{\parskip}{0pt}
\section{Generative AI Use Disclosure}
Large Language Models (LLMs) were used exclusively for language editing, including rephrasing and grammatical refinement, to improve clarity and readability. The LLMs were not involved in the development of ideas, methodology design, experimental procedures, data analysis, or interpretation of results. All scientific content was developed and verified by the authors.
\bibliographystyle{IEEEbib}
\bibliography{references}
\endgroup

\end{document}